\documentclass[structabstract]{aa} 
\newcommand{\kms}{$\mathrm{km\, s^{-1}\, }$}
\usepackage{natbib}
\usepackage{graphicx}

\usepackage{txfonts}
\begin{document}
   \title{Central kinematics of the globular cluster NGC 2808: Upper limit on the mass of an intermediate-mass black hole. \thanks{Based on observations collected at the European Organization for Astronomical Research in the Southern Hemisphere, Chile (083.D-0444).}}
   
   \titlerunning{Central kinematics of the globular cluster NGC 2808}

   \author{N. L\"utzgendorf
          \inst{1}
          \and
          M. Kissler-Patig\inst{1}
          \and
          K. Gebhardt\inst{2}          
          \and
          H. Baumgardt\inst{3}          
          \and
          E. Noyola\inst{4,5}
          \and
          B. Jalali\inst{6}
		  \and
          P. T. de Zeeuw\inst{1,7}
          \and
          N. Neumayer\inst{1}
          }

   \institute{European Southern Observatory (ESO),
              Karl-Schwarzschild-Strasse 2, 85748 Garching, Germany\\
              \email{nluetzge@eso.org}
         \and
			 Astronomy Department, University of Texas at Austin, 
			 Austin, TX 78712, USA 
         \and
			 School of Mathematics and Physics, University of Queensland, 
			 Brisbane, QLD 4072, Australia
         \and
             Instituto de Astronomia, Universidad Nacional Autonoma de Mexico (UNAM), 
             A.P. 70-264, 04510 Mexico
         \and
			 University Observatory, Ludwig Maximilians University, 
			 Munich D-81679, Germany	 
         \and
	         I.Physikalisches Institut, Universit\"at zu K\"oln, 
    	     Z\"ulpicher Str. 77, 50937 K\"oln, Germany
    	 \and
			 Sterrewacht Leiden, Leiden University, 
			 Postbus 9513, 2300 RA Leiden, The Netherlands}

   \date{Received April 10, 2012; accepted April 18, 2012}

 
  \abstract
   {Globular clusters are an excellent laboratory for stellar population and dynamical research. Recent studies have shown that these stellar systems are not as simple as previously assumed. With multiple stellar populations as well as outer rotation and mass segregation they turn out to exhibit high complexity. This includes intermediate-mass black holes which are proposed to sit at the centers of some massive globular clusters. Today's high angular resolution ground based spectrographs allow velocity-dispersion measurements at a spatial resolution comparable to the radius of influence for plausible IMBH masses, and to detect changes in the inner velocity-dispersion profile. Together with high quality photometric data from HST, it is possible to constrain black-hole masses by their kinematic signatures.}
   {We determine the central velocity-dispersion profile of the globular cluster NGC 2808 using VLT/FLAMES spectroscopy. In combination with HST/ACS data our goal is to probe whether this massive cluster hosts an intermediate-mass black hole at its center and constrain the cluster mass to light ratio as well as its total mass.}
   {We derive a velocity-dispersion profile from integral field spectroscopy in the center and Fabry Perot data for larger radii. High resolution HST data are used to obtain the surface brightness profile. Together, these data sets are compared to dynamical models with varying parameters such as mass to light ratio profiles and black-hole masses.}
   {Using analytical Jeans models in combination with variable M/L profiles from N-body simulations we find that the best fit model is a no black hole solution. After applying various Monte Carlo simulations to estimate the uncertainties, we derive an upper limit of the back hole mass of $ M_{BH} < 1 \times 10^4 \ M_{\odot}$ (with $95 \% $ confidence limits) and a global mass-to-light ratio of $M/L_V = (2.1 \pm 0.2) \ M_{\odot}/L_{\odot}$.}
   {}

   \maketitle

   \maketitle
%

\section{Introduction}


Kinematics of globular clusters have long been a field of interest in observational and computational astronomy. In the last years, a new aspect, searching for the signature of central intermediate-mass black holes (IMBHs) was added to the picture. The specific mass range of black holes ($10^2 - 10^5 M_{\odot}$) had not been observed before and for a long time intermediate-mass black holes even thought not to exist. However, recent observations \citep[eg.][]{gebhardt_2000,gerssen_2002,gebhardt_2005,noyola_2008,nora11} have shown that the velocity-dispersion profiles of some globular clusters and dwarf galaxies are consistent with hosting a massive black hole at their center.

Numerical simulations have demonstrated \citep{zwart_2004, gurkan_2004, freitag_2006} that intermediate-mass black holes can form in dense young star clusters by runaway merging. However, taking into account mass loss through stellar winds, \cite{yungelson_2008} found that super-massive stars with initial masses up to $1000 \ M_{\odot}$ reduce to objects less massive than $\sim 150 \ M_{\odot}$ by the end of their lives. An other scenario was presented by \cite{miller_2002} who discussed the formation of intermediate-mass black holes through mergers of stellar-mass black holes in globular clusters when starting with a $M \gtrsim 50 \ M_{\odot}$ seed black hole. Further, they presented scenarios for the capture of clusters by their host galaxies and accretion in the galactic disk in order to explain the observed bright X-ray sources. In addition, remnants of massive Population III stars could have formed intermediate-mass black holes in the early universe \citep{madau_2001}.

If the velocity dispersion - black-hole mass scaling relation observed for supermassive black hole in galaxies \citep[e.g.][]{ferrarese_2000,gebhardt_2000, gultekin_2009} holds at the lower end, intermediate-mass black holes would be expected in systems with velocity dispersions between 10-20 km/s like for globular clusters. Radio and X-ray detection of gas in the central regions is also employed either to provide a black hole mass estimate or an upper limit \citep[e.g.][]{maccarone_2005,ulvestad_2007,bash_2008,cseh_2010, strader_2012}. While the current flux limits of \cite{strader_2012} are impressively low, in order to provide an upper limit on a possible black hole mass, they must make various assumptions about the gas accretion process. Some of the more uncertain assumptions include 1) the distribution of the gas, since a clumpy distribution will lead to time variability which significantly lowers the detection probability, 2) the efficiency of the accretion process, which may be particularly important if these systems are dominated by advection or convection, and 3) uncertainties in translating X-ray fluxes to bolometric fluxes to black hole masses. Thus, non-detection in X-ray and radio can be difficult to interpret in terms of a black hole mass upper limit.

So far the best candidates for hosting an IMBH are the most massive globular clusters in the local group. One of them is $\omega$ Centauri (NGC 5139)  where \citet{noyola_2008, noyola_2010} and \citet{jalali_2012} detect the kinematic signature of a 40~000 $M_{\odot}$ black hole based on radial velocities from integrated light. The results were however challenged by  \cite{vdMA_2010}, who find a lower value of the black-hole mass via HST proper motions. \cite{jalali_2012} perform N-body simulations and show that the current kinematic and light profile with respect to a kinematic center found in Noyola et al. (2010) are consistent with presence of a $5 \times 10^4 \ M_{\odot}$ IMBH assuming a spherical isotropic model for this cluster. Another good candidate is G1 in M31. It is the most massive globular cluster in the Local Group and is found to host a black hole of 20 000 $M_{\odot}$ \cite[eg.][]{gebhardt_2005} by kinematic measurements. In addition  \cite{pooly_2006, kong_2007,ulvestad_2007} detected X-ray and radio emission at its center which is consistent with a black hole of the same mass.

A third good candidate is NGC 6388, a massive globular cluster in our Galaxy. \cite{nora11} detected a rise in its central velocity-dispersion profile which is consistent with a black hole of $\sim$ 20~000 $M_{\odot}$ at its center. The understanding of the formation and evolution of intermediate-mass black holes is crucial for the understanding of the evolution and formation of supermassive black holes. Seed black holes are needed in order to explain the fast formation process of these massive black holes which are observed at very high redshift, i.e. at early times in the Universe \cite{fan_2006}. Intermediate-mass black holes formed in globular clusters and accreted by their host galaxy could be these seeds \cite[eg.][]{ebisuzaki_2001,tanaka_2009}.

We chose to observe the globular cluster NGC 2808 as it shows a variety of interesting features. As known for some globular clusters by now, NGC 2808 has multiple stellar populations. So far only one other cluster ($\omega$ Centauri) shows more than two distinguishable populations \cite[eq.][]{bedin_2004}. As observed by \cite{piotto_2007}, NGC 2808 shows a triple main sequence, which indicates the existence of three sub-populations, all with an age of $\sim 12.5$ Gyr, but with different metallicities. Also, its complex extended horizontal branch morphology \citep{harris_1974,ferraro_1990} shows puzzling discontinuities in the stellar distribution along its blue tail. \cite{maccarone_2008} analyze deep radio observations of NGC 2808 and found no sources detected within the core radius. This places an upper limit on a possible intermediate-mass black hole of $370 - 2100 M_{\odot}$ assuming a uniform gas density and the accretion rate to be different fractions of the Bondi rate. This limit can increase if one assumes even lower Bondi accretion rates or non-uniform gas content in the cluster. 

\cite{noyola_2011} analyze N-body simulations of star clusters with and without central black holes. These reveal that the presence of an IMBH induces a shallow central cusp in the radial density profile. Hence, clusters showing shallow cusps are the best candidates for harboring an IMBH. Further, \cite{miocchi_2007} investigate the effect of an IMBH on horizontal branch morphologies. A central black hole that strips enough stars of their outer envelope during close passages, could be one avenue for producing an extended horizontal branch (EHB). NGC 2808 displays a shallow cusp as well as an EHB, making it an excellent candidate for harboring a central black hole. Given its measured central velocity dispersion of $\sigma = 13.4 \ \mathrm{km}/\mathrm{s}$ \citep{pryor_1993} an extrapolation of the $M_{BH} - \sigma$	 relation predicts a black-hole mass of $\sim 3 \times 10^3 M_{\odot}$. This translates into a radius of influence of $1'' - 2''$ at a distance of $\sim 9.6$ kpc, which would induce clear kinematic signatures inside the $\sim 12''$ core radius. Further main characteristics of NGC 2808 are listed in table \ref{tab_2808}.

This work aims at investigating whether the globular cluster NGC 2808 hosts an intermediate-mass black hole at its center. We first study the light distribution of the cluster. Photometric analysis, including the determination of the cluster center and the measurement of a surface brightness profile, is described in section \ref{phot}. De-projecting this profile gives an estimate of the gravitational potential produced by the visible mass. The next step is to study the dynamics of the cluster. Section \ref{spec} summarizes our FLAMES observations and data reduction and section \ref{kin} describes the analysis of the spectroscopic data. With the resulting velocity-dispersion profile, it is possible to estimate the dynamical mass of the cluster. We then compare the data to Jeans models in section \ref{jeans}. Finally, we summarize our results, list our conclusions and give an outlook for further studies in section \ref{con}.

\begin{table}
\caption{Properties of the globular cluster NGC 2808 from the references: NG=\cite{noyola_2006}, H= \cite{harris_1996} and PM=\cite{pryor_1993}.}             
\label{tab_2808}      
\centering
\begin{tabular}{l l l}
\hline \hline
\noalign{\smallskip}
 Parameter & Value & Reference\\
 \noalign{\smallskip}
\hline
\noalign{\smallskip}
 RA (J2000) & $09\mathrm{h} \ 12\mathrm{m} \ 03\mathrm{s}$  & NG \\ 
 DEC (J2000) & $-64^{\circ}\ 51' \ 49''$ & NG\\ 
Galactic Longitude $l$ & $282.19 °$ &H\\
Galactic Latitude $b$ & $-11.25 °$ &H\\
Distance from the Sun $R_{\mathrm{SUN}}$ & $9.6 \ \mathrm{kpc}$ & H \\
Core Radius $r_c$ & $12.4''$ & NG  \\
Central Concentration $c$ & $1.77$ & H  \\
Heliocentric Radial Velocity V$_r$ & $101.6 \pm 0.7 \ \mathrm{km}/\mathrm{s}$ & H \\ 
Central Velocity Dispersion $\sigma$ & $13.4 \ \mathrm{km}/\mathrm{s}$ & PM \\
Metallicity $[\mathrm{Fe}/\mathrm{H}]$ & $-1.15 \ \mathrm{dex}$ & H \\
Integrated Spectral Type  & F7 &H \\ 
Reddening E(B-V) & $0.22 $ & H \\
Absolute Visual Magnitude $M_{Vt}$ & $-9.39 \ \mathrm{mag}$ & H  \\
\noalign{\smallskip}
\hline 
\end{tabular} 
\end{table}

\section{Photometry}\label{phot}
                                   

The photometric data were taken from the archive of the Hubble Space Telescope (HST). They were obtained with the Wide Field and Planetary Camera 2 (WFPC2) in May 1998 (GO-6804, PI: F. Fusi Pecci) and are composed of a set of two exposures. The deep exposure dataset contains three images each in the filters I (F814W) and V (F555W) with exposure times of 120 and 100 s, respectively. In addition we use a set of two shallow images per filter with exposure times of 3 s in the I filter and 7 s in the V filter obtained in the same run. For both datasets the images cover the cluster center with the central $20''$ on the PC chip. The data were calibrated using the WFPC2-specific calibration algorithm, as retrieved from the European HST-Archive (ST-ECF, Space Telescope European Coordinating Facility\footnote{Based on observations made with the NASA/ESA Hubble Space Telescope, and obtained from the Hubble Legacy Archive, which is a collaboration between the Space Telescope Science Institute (STScI/NASA), the Space Telescope European Coordinating Facility (ST-ECF/ESA) and the Canadian Astronomy Data Centre (CADC/NRC/CSA).}).

\subsection{Color magnitude diagram (CMD) of NGC 2808} \label{sec_cmd}

The CMD is obtained using the programs \textit{daophot II}, \textit{allstar} and \textit{allframe} by P. Stetson, applied to the HST images. For a detailed documentation of these routines, see \cite{stetson_1987}. These programs are especially developed for photometry in crowded fields and are therefore ideally suited for the analysis of globular clusters. For the description of the individual steps we refer to our previous paper \citep{nora11}.

\begin{figure}
   \centering
   \includegraphics[width=0.5\textwidth]{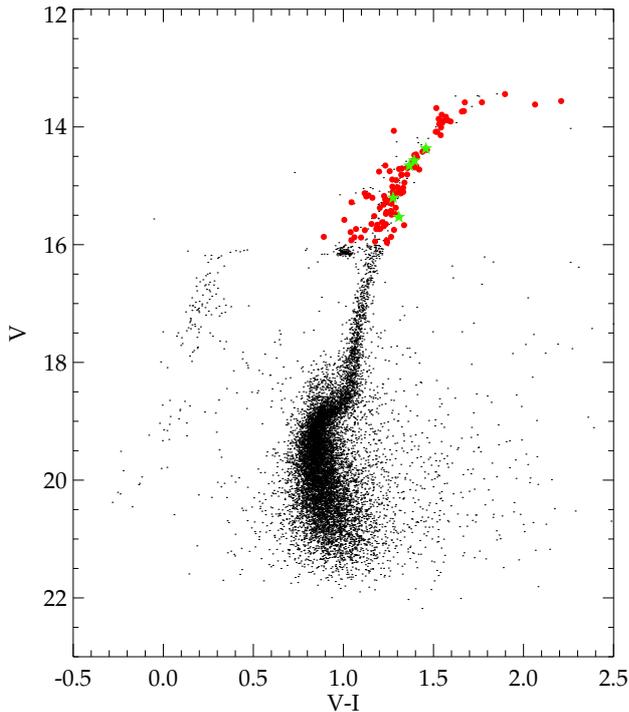}
      \caption{The color-magnitude diagram of NGC 2808. Overplotted are the brightest stars identified in the ARGUS field of view (red circles), and the used template stars (green stars).}
         \label{cmd}
   \end{figure}

We obtain two catalogs with V and I magnitudes from the two datasets with different exposure times. We match and combine the catalogs using the routines \textit{CataXcorr} and \textit{CataComb} in order to obtain a complete star catalog over a wide magnitude range. All  coordinates are transformed to the reference frame of the first I band image of the shallow exposure (\textit{u4fp010br}). Figure \ref{cmd} shows the final CMD with the positions of the brightest stars in the ARGUS pointing and the spectroscopic template stars (see section \ref{spec}) overplotted.

\subsection{Cluster center determination} \label{phot_center}

A crucial step is the determination of the cluster's center. Precise knowledge of the center is important since the shape of the surface brightness and the angular averaged line-of-sight velocity distribution (LOSVD) profiles depend on the position of that center. Using the wrong center typically produces a shallower inner profile. Previous attempts have shown that the larger the core radius of the cluster, the more difficult it is to determine the exact position of the center. For example, the exact position of the center of the globular cluster $\omega$ Centauri is still under debate and differs by up to $12''$ in different analyses \citep[e.g.][]{noyola_2010,AvdM_2010}. In contrast, NGC 6388, with a core radius 10 times smaller than $\omega$ Centauri, turned out to have a well defined center when applying various methods to determine the center of the cluster \citep{ nora11}.

\citet[][hereafter NG06]{noyola_2006} determined the center of NGC 2808 to be at $\alpha = $ 09:12:03.09,  $\delta = -64$:51:48.96  $(J2000)$, with an uncertainty of $0.5''$, by minimizing the standard deviation of star counts in eight segments of a circle. NGC 2808 has a small core radius ($12''$) in comparison to $\omega$ Centauri \citep[NG06;][]{noyola_2010,AvdM_2010}. In order to get an estimate of how accurately the center can be derived, we apply various routines to our catalog.

The field of view of our dataset is very small because we are limited to the $34'' \times 34''$ field of the PC ship. In such a small field of view, it is difficult to determine the center given the large errors arising from the Poisson statistics, i.e. shot noise. Also, if the core is extended, we might not be able to see the stellar concentration decreasing. However, we compensate for this by using different techniques and by estimating the error from their scatter. All techniques are applied for stars brighter than $m_V = 20$ in order to account for the incompleteness effect for faint stars.

The first technique uses isodensity contours as described in \cite{AvdM_2010}. The field of view is divided into boxes of equal size of $100 \times 100$ pixels ($4.6'' \times 4.6''$). This size is the best compromise between having too few stars in each box and therefore large shot noise errors (boxes too small), and having not enough points and therefore a very noisy contour plot (boxes too large). The boxes contain about 100 stars on average. In each box, the stars are counted and the density is derived. The innermost isodensity contours, which were not disturbed by geometrical incompleteness are fitted by ellipses and their central points are determined. From the average of these points and their scatter we determine the central position and its error.

   \begin{figure*}
  \centering
   \includegraphics[width=\textwidth]{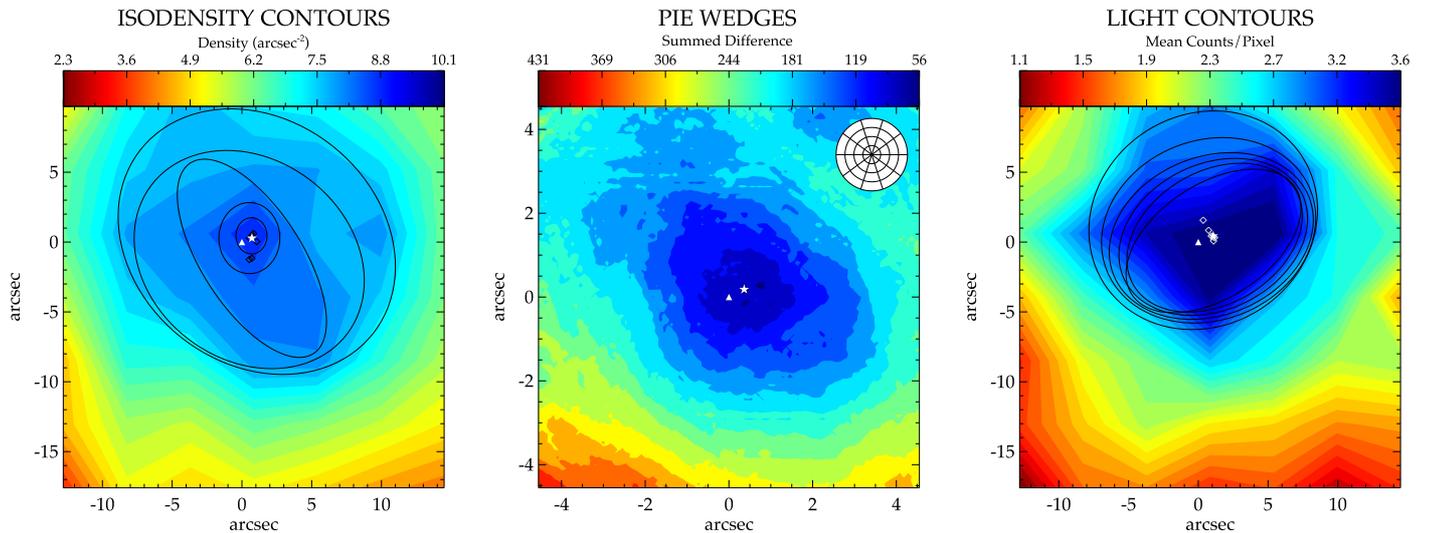}
      \caption{The method to determine the globular cluster center. Left, ellipses fitted to isodensity contours. Middle panel, the contours of the cumulative pie wedges method. The wedges are shown in the upper right of the plot. The right panel shows the contours of the pixel analysis of the mapped image. Contours in blue mark areas of either high stellar density in the left panel, high symmetry (i.e. low differences between the wedges) in the central panel or high light density in the right panel. In every plot the triangle marks the center adopted by NG06 and the white star the newly derived center.}
         \label{center}
   \end{figure*}

The second method is very similar to the one described in \cite{47tuc} and \cite{nora11}. It uses a symmetry argument to determine the cluster center. In our field of view a grid of trial centers is created, using a grid spacing of 2 PC pixels ($\sim 0.1''$). Around each trial center, a circle is traced and divided into wedges as shown in Figure \ref{center}. In order to increase the number of stars, we vary the size of the circle depending on its position on the image. That is, the circle is always as big as the image borders allow it. The stars in each wedge are counted and the numbers compared to the opposite wedge. The differences in the total number of stars between two opposing wedges are summed for all wedge pairs and divided by the area of the circle. The coordinates that minimize the differences define the center of the cluster. We use different numbers of wedges from 4 to 16. We repeat the procedure by rotating the wedges so that their bisector is matching the x-y axis. This method is refined by comparing the cumulative stellar distribution of the stars in the opposing wedges instead of the star counts alone. 

The third and last method is based on the HST/WFPC2 image instead of the star catalog. This method computes the light distribution of the cluster. For this we map out the brightest stars with circular masks in order to prevent a bias towards these stars. We divide the resulting image into boxes of $100 \times 100$ pixel and derive the mean value of the counts per pixel in each box. From this we compute a contour plot and fit ellipses to determine the center.  Despite our attempt to not get biased by the brightest stars, we find that in comparison to the previous two methods, the center obtained with the light distribution is shifted towards a clump of bright stars north-east of the center. However, due to its large errors it does not bias the final center position.

\begin{figure}
  \centering
   \includegraphics[width=0.45\textwidth]{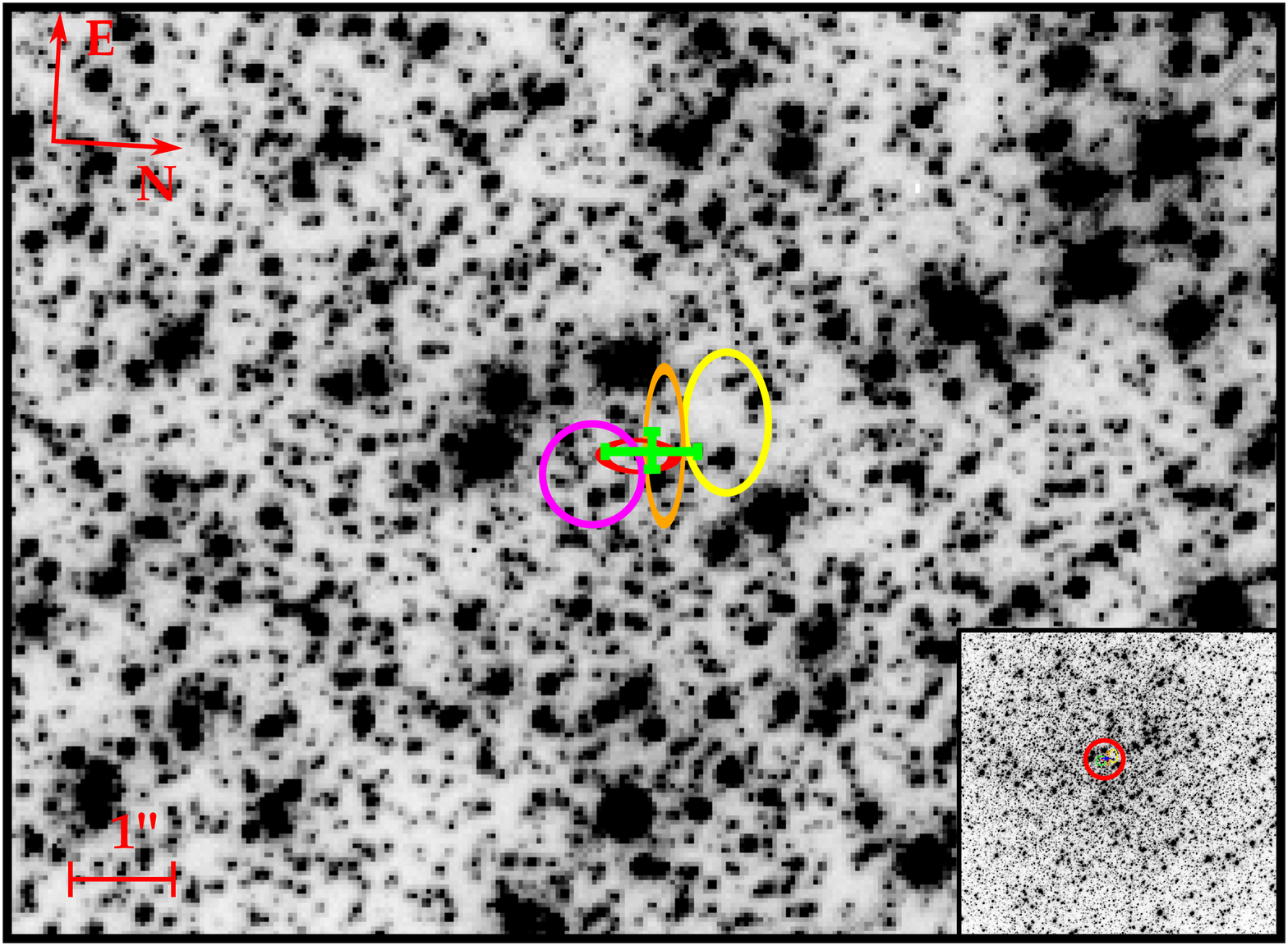}
      \caption{Finding chart for the center of NGC 2808. The magenta circle marks the center of NG06, the red ellipse the results from the pie wedges method, the orange ellipse is for the isodensity contour method and the yellow one shows the position and error of the method measuring the light distribution. The final adopted center and its error bar are displayed in green.}
         \label{find}
   \end{figure}

Figure \ref{center} shows the contour plots of the different methods and Figure \ref{find} presents a finding chart of our final center position. As a final result we obtain
\begin{equation} (x_c,y_c) = (344.4, 441.6) \pm (9.9,3.7) \ \mathrm{pixel} \end{equation} \label{p3}
\begin{eqnarray} \alpha &=& 09:12:03.107,  \ \Delta\alpha = 0.5''  \ \mathrm{(J2000)} \\ \label{p4}
				 \delta &=& -64:51:48.45,  \ \Delta\delta = 0.2'' \end{eqnarray} \label{p4}
which uses as reference image \textit{u4fp010br}. We compared our result to the center obtained by NG06. The two centers are $0.32''$ apart and thus coincide within the error bars of $0.5''$ (as determined by NG06 performing artificial image tests). This center also coincides within $0.3 ''$ with the one derived by \cite{goldsbury_2010} using ellipse fitting applied to the density distribution on ACS/WFC images.

\subsection{Surface brightness profile} \label{phot_sb}

\begin{figure}
  \centering
   \includegraphics[width=0.5\textwidth]{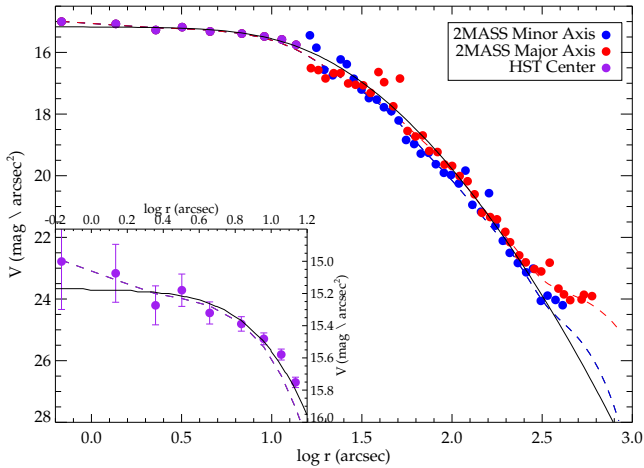}
      \caption{The surface brightness profile of NGC 2808. The red and the blue circles mark the measurements from the 2MASS image along the major and minor axis, respectively, as well as their MGE parametrization (dashed lines). The profile obtained from the HST star catalog is shown in purple. Overplotted is the profile obtained by \cite{trager_1995} with a solid black line.}
         \label{sb}
   \end{figure}

The surface brightness (SB) profile is required as an input for the Jeans models described in the following section. As in \cite{nora11} we use a simple method of star counts in combination with an integrated light measurement from the WFPC2 image to derive the SB profile. For our analysis we do not use stars brighter than $m_V = 17$ to avoid contamination by very bright stars in the center. The fluxes of all stars brighter than $m_V = 20$ are summed in radial bins around the center and divided by the area of the bin. In addition, the integrated light for stars fainter than $m_V = 20$ is measured directly from the HST image. Using the same radial bins as in the star count method, we measure the statistical distribution of counts per pixel excluding regions containing stars with $m_V < 20$ by mapping out these stars in the image using a circular mask with a radius of $0.3 ''$ (6 HST pixels). We use a robust bi-weight estimator to derive the mean counts-per-pixel for every bin. Finally, the flux per pixel is transformed into magnitudes per square arcseconds and added to the star count profile. Due to the small field of view of the PC chip we are only measuring points inside a radius of $10 ''$. The errors of our profile are obtained by Poisson statistics of the number of stars in each bin. With a linear fit to the innermost points ($r < 10''$) we derive a slope of the surface luminosity density $I(r) \varpropto r^{\, \alpha}$ of $\alpha = -0.16 \pm 0.08$. This value is steeper (but consistent within the errors) than the slope of $\alpha =-0.06 \pm 0.07$ derived by NG06. The final inner profile is listed in Table \ref{tab_sb} and shown in Figure \ref{sb}.

For the outer regions we use the profile obtained by \cite{trager_1995} for our spherical models and a two-dimensional profile obtained from ground based (2MASS) images (see Section \ref{axis}) for the axisymmetric models. The two 2MASS images in J-band which cover the entire cluster were received from the public archive and combined to a single pointing. Using routines provided in the package of the anisotropic Jeans models we obtain a two-dimensional surface brightness profile by fitting isophotes to the image. The final profile is constructed by combining the inner profile obtained with the HST with the outer profile from the 2MASS image. Since the two images are taken in different bands, it is necessary to scale the images to a common flux. For this purpose we simply scale the data points of the 2MASS image to the HST profile. Here we neglect stellar population effects and assume a constant color within the cluster. Figure \ref{sb} shows the final combined profile with the minor and major axis of the two-dimensional profile colored in blue and red, respectively.

\section{Spectroscopy}\label{spec}

The spectroscopic data were obtained with the GIRAFFE spectrograph of the FLAMES (Fiber Large Array Multi Element Spectrograph) instrument at the Very Large Telescope (VLT) using the ARGUS mode (Large Integral Field Unit). The observations were performed during two nights (2010-05-05/06). The ARGUS unit was set to the 1 : 1.67 magnification scale (pixel size: $0.52 ''$, $14 \times 22$ pixel array) and pointed to six different positions, each of them containing three exposures of 600s with $0.5''$ dithering to cover the entire core radius. The position angle of the integral field unit remained at 0 degrees (long axis parallel to the north-south axis) during the entire observation.

\begin{table}
\caption{The derived surface brightness profile in the V-band. $\Delta V_{h}$ and $\Delta V_{l}$ are the high and low values of the errors, respectively}
\label{tab_sb}      
\centering
\begin{tabular}{c c c c}
\hline \hline
\noalign{\smallskip}
  
$\log r$ 	& $V$  								&$\Delta V_{l}$ 					& $\Delta V_{h}$ \\ 

[arcsec] 			& $[\mbox{mag} / \mbox{arcsec}^2]$ 	&$ [\mbox{mag} / \mbox{arcsec}^2]$ 	& $[\mbox{mag} / \mbox{arcsec}^2]$ \\ 
\noalign{\smallskip}
\hline
\noalign{\smallskip}

$-0.17$ & $15.00$ & $0.30$ & $0.24$ \\
$0.14$ & $15.07$ & $0.18$ & $0.16$ \\
$0.36$ & $15.27$ & $0.12$ & $0.11$ \\
$0.50$ & $15.18$ & $0.10$ & $0.09$ \\
$0.66$ & $15.32$ & $0.07$ & $0.06$ \\
$0.83$ & $15.39$ & $0.04$ & $0.04$ \\
$0.96$ & $15.48$ & $0.04$ & $0.04$ \\
$1.06$ & $15.58$ & $0.03$ & $0.03$ \\
$1.14$ & $15.75$ & $0.03$ & $0.03$ \\
      
\noalign{\smallskip}
\hline
\end{tabular}
\end{table}

The kinematics are obtained from the analysis of the Calcium Triplet ($\sim 850 \, \mathrm{nm}$), which is a strong feature in the spectra. The expected velocity dispersions lie in the range 5-20 \kms and had to be measured with an accuracy of 1-2 \kms. This implied using a spectral resolution around $10~000$, available in the low spectral resolution mode set-up LR8 ($820-940 \, \mathrm{nm}, \, \mathrm{R} = 10~400$).


We reduce the spectroscopic data with the GIRAFFE pipeline programmed by the European Southern Observatory (ESO). This pipeline consists of five recipes (\textit{gimasterbias, gimasterdark, gimasterflat, giwavecalibration, giscience}) which are described in \cite{nora11}. From the input observations, the final routine \textit{giscience} produces a reduced science frame as well as the extracted and rebinned spectra frame. At the end, the recipe also produces a reconstructed image of the respective field of view of the ARGUS observations.

For sky subtraction, we use the program developed by Mike Irwin and described in \cite{battaglia_2008}. The program combines all 14 sky fibers using a 3-sigma clipping algorithm and computes an average sky spectrum. It splits the continuum and the line components for both the averaged sky spectrum and the object spectrum, using a combination of median and boxcar. The sky-line mask and the line-only object spectra are compared finding the optimum scale factor for the sky spectrum and the sky-lines are subtracted from the object spectra. The continuum is added back to the object spectra after subtracting the sky continuum by the same scaling factor.

As a next step we use the program LA-Cosmic developed by \cite{Lacos} to remove the cosmic rays from our spectra. In order to avoid bright stars dominating the averaged spectra when they are combined, we apply a normalization to the spectra by fitting a spline to the continuum and divide the spectra by it. 
                                   

\section{Kinematics} \label{kin}
                                   
In this section we describe how we compute the velocity map in order to check for peculiar kinematic signals. Further, we measure the velocity-dispersion profile which is used to fit analytic models, described in the next section.

\subsection{Velocity map} \label{vel_map}

To construct the velocity map, we use the relative shifts of the pointings to stitch them together and create a catalog with each spectrum correlated to one position in the field of view. The resulting catalog of spectra and their coordinates allow us to combine spectra in different bins. The combined pointing contains 54 $\times$ 44 spaxels and is cross shaped.  We also reconstruct the ARGUS pointing on top of the HST image in order to obtain a direct connection between spaxel positions and our star catalog.

   \begin{figure*}
  \centering
   \includegraphics[width=\textwidth]{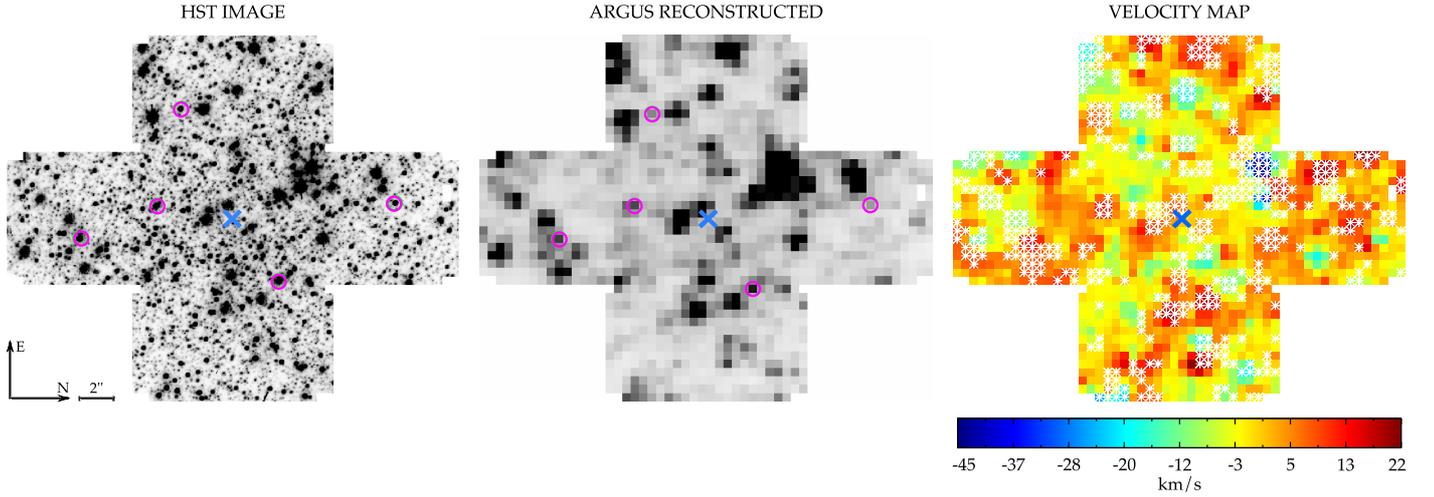}
      \caption{The velocity map of NGC 2808. Shown are the ARGUS field of view reconstructed on the WFPC2 image (left panel), the combined reconstructed ARGUS pointings (middle panel), and the resulting velocity map. Magenta circles mark the template stars used to derive the kinematics and the blue cross marks the center of the cluster. The white crosses on the velocity map mark the spaxel which are not used in deriving the velocity-dispersion profile.}
         \label{vel}
   \end{figure*}

Using the penalized pixel-fitting (pPXF) program developed by \cite{cappellari_2004}, a velocity is derived for each spaxel. In regions of overlap the spectra are first averaged before being analyzed. Figure \ref{vel} shows a) the field of view of the six ARGUS pointings on the HST image, b) the reconstructed and combined ARGUS image and c) the corresponding velocity map. For every ARGUS pointing, an individual velocity map is derived. The velocities of overlapping spaxels are in agreement with each other to 1 \kms, which is expected as discussed in \cite{nora11}.

In order to test individual spaxels for shot noise, we apply the same routines as described in \cite{nora11}. At every position of a star in the catalog, a two dimensional Gaussian is modeled with a standard deviation set to the seeing of the ground based observations ($\mathrm{FWHM}=0.8''$) and scaled to the total flux of the star. We measure the absolute amount and fraction of light that each star contributes to the surrounding spaxels. For each spaxel we then have the following information: a) how many stars contribute to the light of that spaxel, and b) which fraction of the total light is contributed by each star. The test shows that NGC 2808 is less concentrated in the center than NGC 6388 (in fact NGC 6388 has 1.5 times the central stellar density of NGC 2808) and therefore fewer stars contribute to individual spaxel. Also, more spaxels are dominated by a single star by 60~\% or higher. As in the case of NGC 6388, we map out spaxels in which either less than 10 stars contribute to the light or a single star contributes more than 60~\% of the light. This leaves us with 1080 spectra out of the 1514 spectra sample.


The velocity template used in the kinematic analysis is important to consider carefully given the strong changes in the intrinsic line widths in a globular cluster. In order to find an optimal velocity template we first collect all stars which dominate a spaxel by more than 80\% via the shot noise routine described above. We plot these stars on our CMD in order to check for non-cluster members. We choose the five faintest stars and combine them after shifting them to the same velocity. We also try kinematic fits with individual stars with high signal-to-noise from the upper giant branch. We do not find as good a fit to the integrated light using the stars from the upper giant branch as templates; this result is expected since the integrated light comes primarily from the fainter stars. The kinematics, however, are similar using both templates.

The positions of the template stars are marked in the HST image (Figure 5, left panel) with magenta circles and in the CMD (Figure 1) with green stars. We also identify the brightest stars from the pointing in the CMD to make sure that none of the dominating stars is a foreground star (see Figure 1, red dots). In order to derive an absolute velocity scale, the line shifts of the template are measured by fitting a Gaussian to each Ca-triplet line of the template spectrum, and deriving the centroid. This is compared with the values of the Calcium Triplet in a rest frame and the average shift is calculated. The derived radial velocity is transformed to the heliocentric reference frame. This results in a template velocity of $v_{\rm{temp}} = (122.3 \pm 2.3)$ \kms.

As a conspicuous feature in the velocity map we recognize two blue spots in the upper right, indicating two stars with high approaching velocities. Considering the velocity scale, also plotted in Figure \ref{vel}, these features refer to velocities of $-40$ and $-45$ \kms relative to the cluster, respectively. This corresponds to 3.1 and 3.5 times the velocity dispersion \cite[if one assumes the value of][]{pryor_1993}. High velocity stars have been discovered in only a few globular clusters up to now \citep{gunn_1979, meylan_1991} and require a detailed analysis in terms of membership, the tails of the velocity distribution, and ejection mechanisms. We discuss the two high-velocity stars in a separate paper (L\"utzgendorf et al., 2012 in preparation).

\subsection{Inner velocity-dispersion profile}

For the radial velocity-dispersion profile of NGC 2808, we bin the spectra in the following way. The pointing is divided into five independent angular bins, with radii of 2, 5, 10, 20 and 28 ARGUS spaxels corresponding to $1.0'', 2.6'', 5.2'', 10.4''$ and $15.6''$ ($\sim 0.05, 0.12, 0.24, 0.48, 0.68$ pc). We try different combinations of bins and bin distances as well as overlapping bins and find no change in the global shape of the profile. In each bin, all spectra of all exposures are combined with a sigma clipping algorithm to remove any remaining cosmic rays. Velocity and velocity-dispersion profiles are computed by applying pPXF to the binned spectra using the same template as for the velocity map. We compare the results of pPXF with a non-parametric fit. We obtain the best agreement of these two methods by fitting four moments of the Gauss-Hermite parametrization. The final parameters ($\rm V, \sigma, h_3, h_4$) are transformed to the "true" moments of the LOSVD ($\rm \tilde{V}, \tilde{\sigma}, \xi_3, \xi_4$) by applying equations (17) of \cite{vdm_1993}.

We estimate the radial velocity of the cluster in a heliocentric reference frame and the effective velocity dispersion $\sigma_e$. We combine all spectra in the pointing and measure the velocity relative to the velocity of the template. This value is corrected for the motion of the template and the heliocentric velocity and results in a value of  $\rm V_r = (104.3 \pm 2.3)$ \kms which agrees within the errors with the value from \cite{harris_1996} $\rm V_r = (101.6 \pm 0.7)$ \kms. The effective velocity dispersion is derived using equation (1) in \cite{nora11} and results in $\sigma_e= (13.4 \pm 0.2)$ \kms. This is in very good agreement with the central velocity dispersion $\sigma_c = (13.4 \pm 2.6)$ \kms of \cite{pryor_1993}.

\begin{figure}
  \centering
   \includegraphics[width=0.5\textwidth]{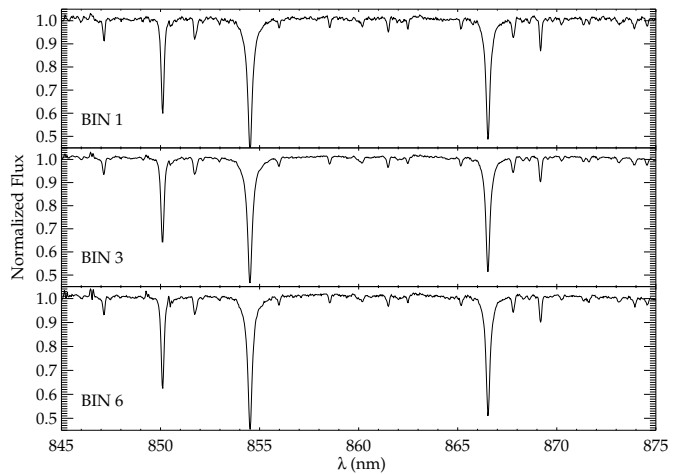}
      \caption{The combined spectra for the bins 1, 3 and 6 from which the kinematic measurements are taken. }
         \label{spec}
\end{figure}

\begin{table*}        
\centering
\caption{The kinematics of NGC 2808 obtained from the VLT/FLAMES data.} 
\label{tab_moments}      
\begin{tabular}{cccccccccccccccccccc}
\hline \hline
\noalign{\smallskip}
  
$\log r$ 	& \multicolumn{3}{c}{$\tilde{\rm V}$} &\multicolumn{3}{c}{$\tilde{\sigma}$} &\multicolumn{3}{c}{${\rm V}$} &\multicolumn{3}{c}{${\sigma}$} &\multicolumn{3}{c}{h$_3$} &\multicolumn{3}{c}{h$_4$}&  S/N  \\ 

[arcsec] 			& \multicolumn{3}{c}{$[\mbox{km} / \mbox{s}]$} 	&\multicolumn{3}{c}{$[\mbox{km} / \mbox{s}]$} & \multicolumn{3}{c}{$[\mbox{km} / \mbox{s}]$} 	&\multicolumn{3}{c}{$[\mbox{km} / \mbox{s}]$}&\multicolumn{3}{c}{}&\multicolumn{3}{c}{} &  \\ 
\noalign{\smallskip}
\hline
\noalign{\smallskip}

\multicolumn{20}{c}{FLAMES MEASUREMENTS} \\
\noalign{\smallskip}
$1.09$  & $0.6 $& $\pm$ &$0.3$ & $11.4$& $\pm$ &$2.5$& $0.3 $& $\pm$ &$0.3$ & $13.0$& $\pm$ &$0.6$ & $0.00$& $\pm$ &$0.01$ & $-0.15$& $\pm$ &$0.06$ &$  128$ \\
$2.60$  & $-0.0$& $\pm$ &$0.3$ & $13.0$& $\pm$ &$1.3$& $-0.5$& $\pm$ &$0.3$ & $14.7$& $\pm$ &$0.6$ & $0.02$& $\pm$ &$0.00$ & $-0.16$& $\pm$ &$0.07$ &$  165$ \\
$5.20$  & $-1.6$& $\pm$ &$2.9$ & $13.8$& $\pm$ &$0.7$& $-2.4$& $\pm$ &$2.9$ & $15.5$& $\pm$ &$0.4$ & $0.04$& $\pm$ &$0.22$ & $-0.12$& $\pm$ &$0.03$ &$  188$ \\
$7.80$  & $-0.1$& $\pm$ &$0.4$ & $13.6$& $\pm$ &$0.5$& $-0.5$& $\pm$ &$0.4$ & $15.4$& $\pm$ &$0.6$ & $0.01$& $\pm$ &$0.00$ & $-0.14$& $\pm$ &$0.06$ &$  184$ \\
$10.40$ & $0.9 $& $\pm$ &$0.4$ & $13.9$& $\pm$ &$0.5$& $0.4$& $\pm$ &$0.4$ & $15.7$& $\pm$ &$0.5$ & $0.02$& $\pm$ &$0.02$ & $-0.14$& $\pm$ &$0.00$ &$  157$ \\
$14.56$ & $0.9 $& $\pm$ &$0.4$ & $13.1$& $\pm$ &$0.6$& $0.5$& $\pm$ &$0.4$ & $14.8$& $\pm$ &$0.7$ & $0.02$& $\pm$ &$0.00$ & $-0.17$& $\pm$ &$0.09$ &$  154$ \\
\noalign{\smallskip}
\multicolumn{20}{c}{FABRY-PEROT MEASUREMENTS} \\
\noalign{\smallskip}
$57.42$ & $-0.2$& $\pm$ &$0.4$ & $10.7$& $\pm$ &$0.3$&&&&&&&&&&&&& \\
$89.27$ &  $-1.4$& $\pm$ &$1.3$ & $9.4$& $\pm$ &$0.9$&&&&&&&&&&&&& \\
$108.02$ &  $1.5$& $\pm$ &$0.9$ & $8.2$& $\pm$ &$0.7$&&&&&&&&&&&&& \\
$157.58$ &  $1.3$& $\pm$ &$1.0$ & $8.1$& $\pm$ &$0.7$&&&&&&&&&&&&& \\

\noalign{\smallskip}
\hline
\end{tabular}
\end{table*}
For the error estimation we run Monte Carlo simulations for each bin. From the routine described in section \ref{vel_map}, we know how many stars contribute to what amount to each spaxel, and how many spaxels are added up in each bin. Each of the stars in one bin is assigned a velocity chosen from a Gaussian velocity distribution with a fixed dispersion of 10 \kms. Using our template spectrum we shift the spectra by their velocity and weight them according to their contribution before combining them in to one spaxel. The resulting spaxels are  normalized, combined and the kinematics measured with pPXF (as for the original data). After 1000 realizations for each bin, we obtain the shot noise errors from the spread of the measured velocity dispersions. The errors for the velocity are derived by applying Monte Carlo simulations to the spectrum itself. This is done by repeating the measurement for 100 different realizations, adding noise to the original spectra \citep[see ][section 3.4]{cappellari_2004}.

\begin{figure*}
  \centering
  \includegraphics[width=\textwidth]{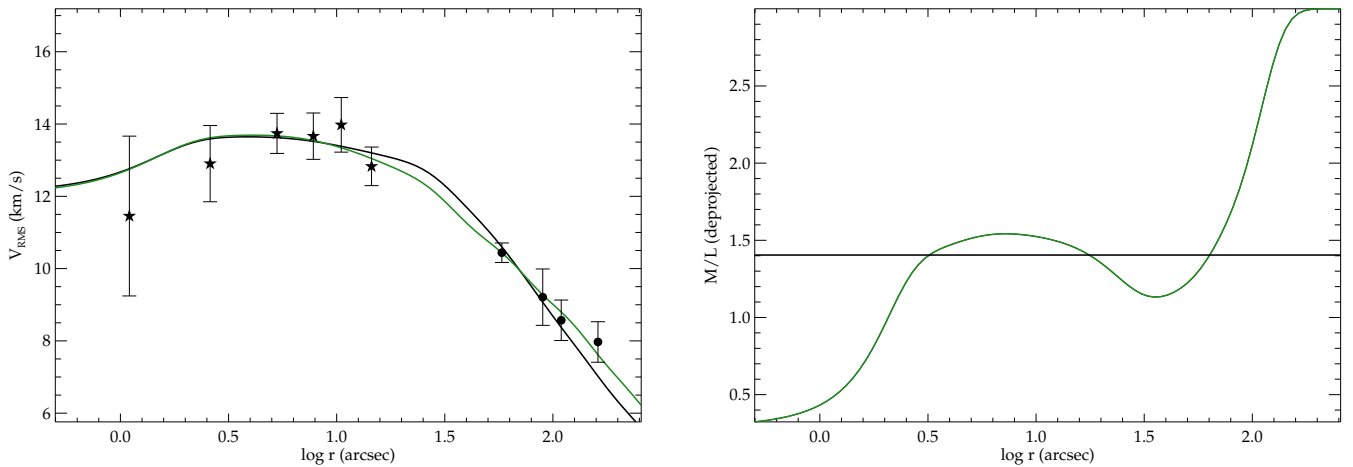}
     \caption{Isotropic spherical Jeans models compared to the kinematic data of NGC 2808. Stars correspond to the inner kinematic profile obtained with integral-field spectroscopy and bullets represent the data obtained with the Fabry-Perot instrument. Shown are a model with a constant $M/L_V$ over the entire radius of the cluster (black line) as well as the model where we fit an $M/L_V$ profile (green line). The resulting $M/L_V$ profile is shown in the right panel together with the constant value of the previous model.}
        \label{fig:mod_spher}
\end{figure*}

The resulting profile is displayed in Figure \ref{fig:mod_spher}. The innermost point drops down to a dispersion of 11.4 \kms, which is lower than the outermost point of the IFU data, but it is also severely affected by shot noise, as seen by its large error bar. In Table \ref{tab_moments}, we record the results of the kinematic measurements. The first column lists the radii of the bins. The following columns show the central velocities of each bin in the reference frame of the cluster, the corrected velocity dispersion $\tilde{\sigma}$ (V$_{RMS}$), as well as the parameters from the Gauss-Hermite parameterization V, $\sigma$, h$_3$ and h$_4$.

\subsection{Outer kinematics}

In addition to the inner kinematics we use the dataset of Gebhardt et al. (2012, in preparation) for larger radii. This data was obtained in three epochs (1995, 1997, 1998) with the Rutgers Fabry Perot on the Blanco 4-m telescope at Cerro Tololo Inter-American Observatory (CTIO). The data set contains over 3600 velocities of individual stars out to $4'$. For reduction and more detailed information we refer to Gebhardt et al. (2012, in preparation).

The Fabry-Perot velocities come from a very similar set of observations and reductions as presented in \cite{gebhardt_1997}. These spectra are centered on a small region around the H-alpha absorption line, with absolute velocity calibration derived from comparison with published radial velocities. We exclude the Fabry-Perot data in the central regions where crowding is important; whenever measuring radial velocities of individual stars in clusters, one must be careful to limit contamination by other cluster members, otherwise potentially biasing the velocity dispersion measurement.

A closer look at the outer velocities shows a clear rotation at larger radii. We derive the rotation velocity and orientation in the following way: Different radial bins of $10", 30", 50", 70"$ and $90"$ are divided into 12 angular bins and the velocity of each of these wedges is derived. The rotation curve (the velocity as a function of position angle) is fitted by a sine function and orientation as well as rotation velocity is extracted for each of the bins. We obtain an average orientation angle of of maximum positive rotation of $\theta = 132^{\circ} \pm 9^{\circ}$ (measured from North to East). The rotation velocity increases from no rotation within the core radius to $\sim 5$ \kms in the outskirts of the cluster. 

In order to obtain the velocity-dispersion profile, we apply the maximum likelihood method introduced by \cite{pryor_1993}. The iterative scheme used to solve the maximum likelihood equations is very similar to that used by \cite{gunn_1979} to fit the velocity scale parameter of King models. This approach is necessary especially when the uncertainties of the individual velocities are different. The data is divided into radial bins and the velocity dispersion derived for each bin. Due to crowding in the cluster center, fainter stars get contaminated from background light which biases their measured velocities towards the cluster mean velocity. This results in a lower velocity dispersion for points measured in the central regions of the cluster. For this reason we only used measurements for radii larger than $50 ''$ ($\sim r_h$). The resulting velocity-dispersion profile is shown in Figure \ref{fig:mod_spher} together with the points of the IFU measurements and the best fit Jeans model (see section \ref{jeans}).

\section{Dynamical Models} \label{jeans}

After having extracted the velocity dispersion profile over a large radial range, the next step is to compare this data, together with the photometric profiles, to dynamical models. This section describes the different types of models which we compare to our data. We start with a spherical isotropic Jeans model, increase the complexity of the models to axisymmetric models, and radial varying $M/L_V$ profiles.

\subsection{Isotropic spherical Jeans models}

Assuming spherical symmetry is a good approximation for most globular clusters and a valid first order assumption for NGC 2808. To compute the models, we use the Jeans Anisotropic multi-Gaussian expansion (JAM) dynamical models implementation for stellar kinematics of spherical and axisymmetric galaxies developed by \cite{cappellari_2002,cappellari_2008} \footnote{Available at http://www-astro.physics.ox.ac.uk/$\sim$mxc/idl}. The routines take a one dimensional surface brightness profile as an input and use the multi-Gaussian expansion (MGE) method \cite{emsellem_1994} to fit and deprojects the light profile. This is then fed into the spherical Jeans equations and a second-moment profile (V$_{\rm{RMS}} = \sqrt{(\sigma^2 + V_{rot}^2)}$, hereafter referred to as velocity-dispersion profile) is computed. The modeled velocity-dispersion profile is scaled by a constant factor to fit the kinematic data. This scaling factor is adopted as the global $M/L_V$ value. In Figure \ref{fig:mod_spher}, the black line in the left panel represents the spherical Jeans model with a $M/L_V = 1.4$.

The comparison of the spherical model with our data in Figure \ref{fig:mod_spher} shows already good agreement. However, a constant $M/L_V$ profile over a larger radial range is not a good assumption for a globular cluster. N-body simulations have shown that the $M/L_V$ increases for larger radii due to mass segregation and migration of low-mass stars towards the outskirts of the cluster. The underestimation of the model for the outer points might be caused by this effect. Therefore, it is necessary to allow for a varying $M/L_V$ profile in the models.  For this, we apply two methods: the first one is to let the model fit the $M/L$ profile to the data. This can be done by multiplying each of the MGE Gaussians of the fitted surface brightness profile with different factors $p$ \citep{williams_2009}. These factors are varied over a physical range ($p \in \{0,3\}$). For every combination, the Jeans model is computed and the quality of the fit calculated via a least square statistic. With this technique, we find the $M/L_V$ profile which best reproduces the data. The disadvantage of this method is the degeneracy of the problem. There are many combinations of the factors $p$ that return similar quality of the fit to the kinematic data. In addition, without constraining the parameters of the $M/L_V$ profile to a physical limit, the best fit would favor unrealistic high values in the outer regions to fit the flat velocity-dispersion profile beyond $100''$. The resulting $M/L_V$ profile is shown in the right panel of Figure \ref{fig:mod_spher} as a green line.

The second method takes an already existing $M/L_V$ profile obtained from N-body simulations and is explained in the next Section. 

\subsection{Isotropic axisymmetric Jeans models} \label{axis}

The Fabry-Perot dataset shows a clear rotation in the outer regions with rotation velocities up to 5 \kms. The shape of NGC 2808 reveals the rotation character of that cluster with a flattening of $\epsilon= 1 - b/a = 0.12$ \citep{white_1987}. For that reason we compute axisymmetric Jeans models in addition to the spherical models described in the previous section. These models assume an axial symmetry of the velocity ellipsoids rather than a spherical symmetry and are ideally suited for rotating systems. By applying the general axisymmetric Jeans equations (3) and (4) in \cite{cappellari_2008}, the Jeans models provide good descriptions of the two-dimensional shape of the velocity (V) and its velocity-dispersion (V$_{\rm{RMS}}$), once a surface brightness profile is given.

\begin{figure*}
  \centering
  \includegraphics[width=\textwidth]{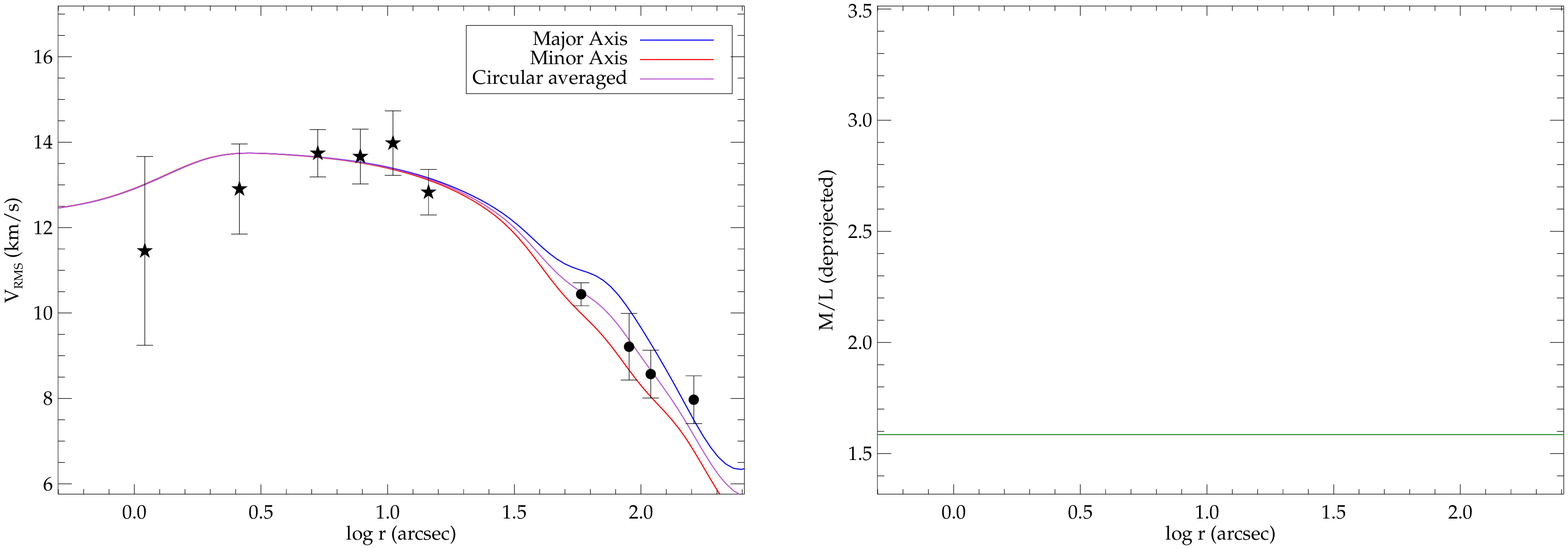}
   \includegraphics[width=\textwidth]{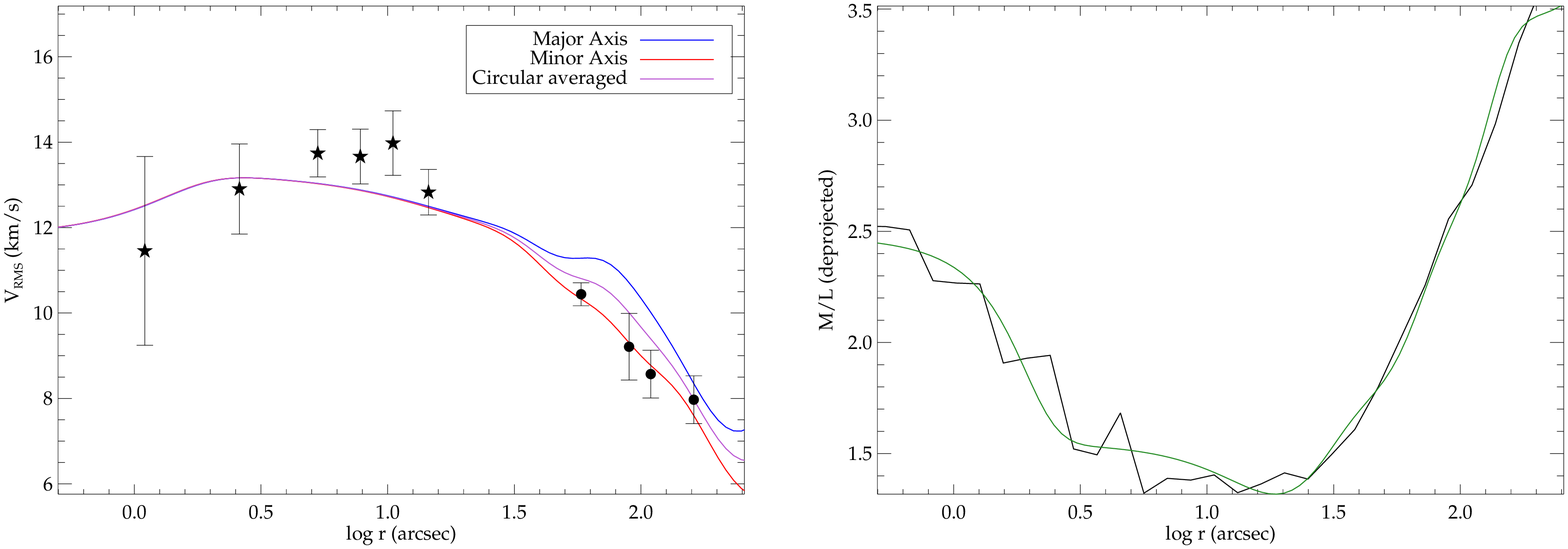}
     \caption{Axisymmetric Jeans models for NGC 2808 for a constant (upper panel) and a varying $M/L_V$ profile (lower panel). The used $M/L_V$ profiles/values are shown in the right panels. The black line marks the inserted profile from N-body simulations and the green line its parametrisations with the MGE Gaussians.}
        \label{fig:mod_axi}
\end{figure*}

\begin{figure*}
  \centering
  \includegraphics[width=\textwidth]{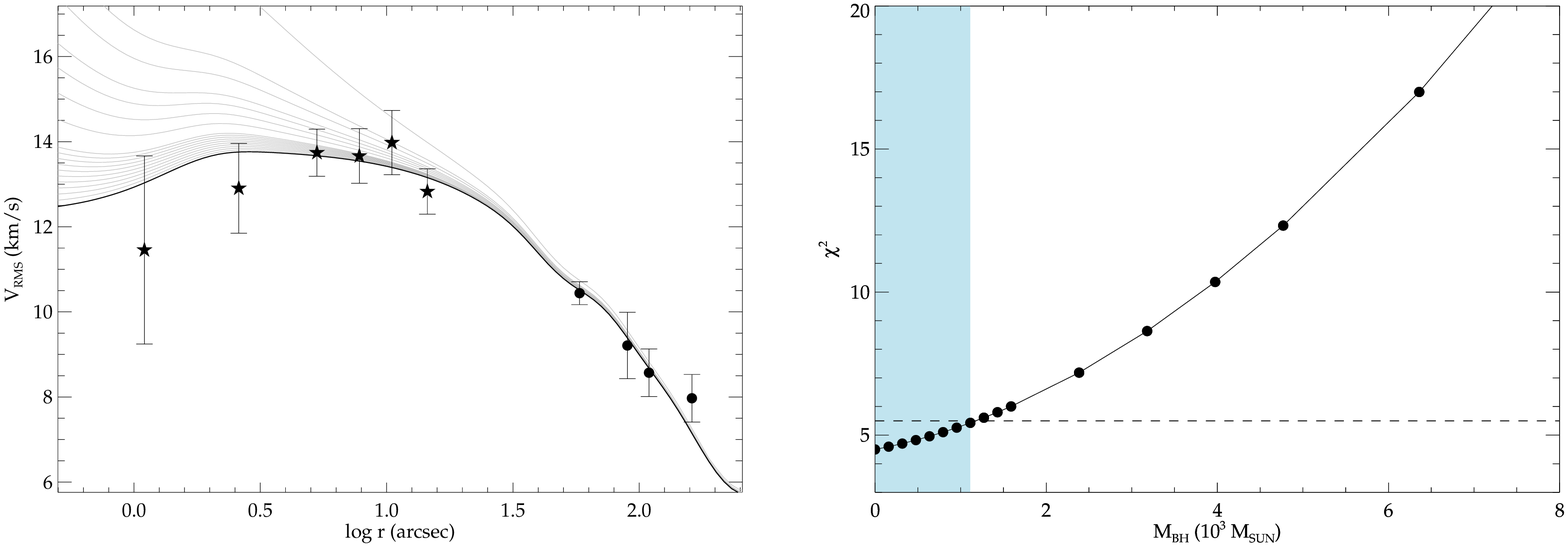}
   \includegraphics[width=\textwidth]{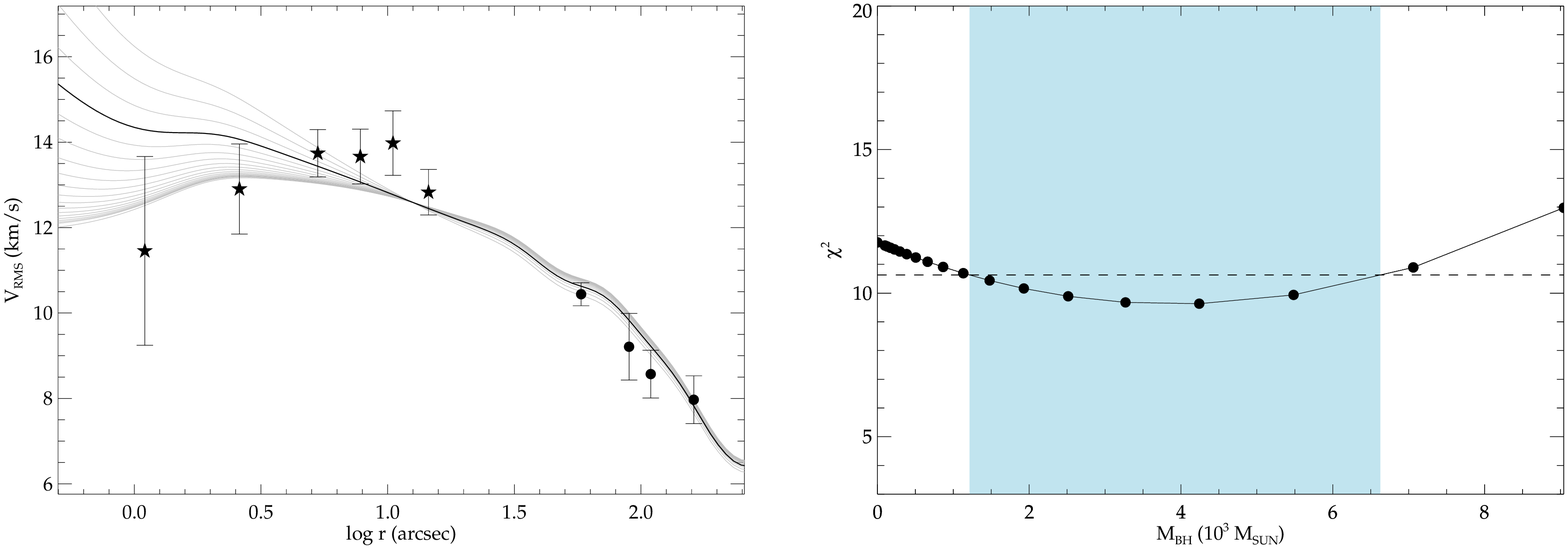}
     \caption{Axisymmetric Jeans models with different black-hole masses for NGC 2808. Upper panel: The model with the constant $M/L_V$ (same as in Figure \ref{fig:mod_axi}). Lower panel: The model with an $M/L_V$ profile. The $/chi^2$ values as a function of black-hole mass are shown in the right panels. Blue shaded areas mark the black-hole mass range where $\Delta \chi^2 < 1$, thus the $1 \sigma$ error. }
        \label{fig:mod_bh}
\end{figure*}

Instead of using an one-dimensional surface brightness profile the axisymetric models (also present in the JAM package) perform a two-dimensional fit to the surface brightness obtained from an image of the object. For globular clusters, determining a 2D surface brightness profile is challenging since the individual stars are resolved and the routine has difficulties in fitting isophotes to a discontinuous image. Images with a lower resolution are better suited for this purpose. We use a combination of a ground based image in J-band taken from publicly available data of 2MASS and of our inner one-dimensional surface brightness profile obtained from the HST image (see Section \ref{phot_sb}). Since the high resolution of the HST does not allow a two-dimensional fit on the image itself and because we do not measure any rotation within the core radius, we assume spherical symmetry inside a radius of $r \sim 15'' $(the size of the combined ARGUS field of view). This assumption is further supported by the results of \cite{nora11} where we found that any anisotropy in a globular cluster is smoothed out within a few relaxation times inside the core radius of the cluster.

For the outer region of the cluster the routine \textit{find\_ galaxy.pro}, which is also included in the JAM package, measures an ellipticity of $\epsilon = 1-b/a = 0.11 $ and an orientation of $\theta = 104^{\circ}$. This orientation differs from the one that we obtain from the kinematic studies ($\theta = 132^{\circ}$) and from the results of \cite{white_1987} of $\theta \sim 121^{\circ}$. Because of the large uncertainties of the photometric measurement (due to shot noise of single stars), we decide to adopt the orientation from the kinematics as our final result $\theta = 132^{\circ} \pm 9^{\circ}$. The next step is to determine the two-dimensional surface brightness profile from the 2MASS image. This is done by the routine \textit{sectors\_ photometry.pro}. The routine performs photometry of an image along sectors equally spaced in angle. The result is converted into magnitudes per squarearcsecond and combined with the inner surface brightness profile obtained with the HST.

The JAM code calculates a two dimensional V$_{\rm{RMS}}$ map by using the surface brightness profile. To compare with our one dimensional kinematic profile we extract the profile along the major axis and the minor axis. In addition, we derive a third profile by averaging the two-dimensional V$_{\rm{RMS}}$ map over concentric radial bins and call it the averaged profile. Figure \ref{fig:mod_axi} shows these three profiles on top of the data. The models differ only in the outer region since the inner part is assumed to be spherical. 

In order to use the most realistic $M/L_V$ profile, we use an $M/L_V$ profile obtained by running N-body simulations and comparing them to our data points using the N-body code NBODY6 \citep{aarseth_2010}. Starting from a \cite{king_1962} model the simulations run with varying initial central concentrations $c$, half-mass radii $r_h$, and IMBH masses with $N = 50~000$ particles. We use a metallicity of $[Fe/H]=-1.18$ and model a stellar evolution according to the stellar evolution routines of \cite{hurley_2000}. The simulations start with stars distributed according to a \cite{kroupa_2001} mass function in the mass range $0.1 < m / M_{\odot} < 100$. Binaries can form during the evolution of the cluster, but are not included primordially. For more informations about these simulations we refer to \cite{mcnamara_2012} and \cite{baumgardt_2003}. We use a grid of computed models with different initial conditions and find the model which fits the observed kinematic and light profile the best. The $M/L$ of this model is then computed as a function of radius by only using the brightest stars. This allows an independent determination of the $M/L$ profile of NGC 2808. This is fed into the Jeans model, which parametrizes the input $M/L$ profile with an MGE fit and applies it to the resulting velocity-dispersion profile. The lower right panel of Figure \ref{fig:mod_axi} shows the used $M/L_V$ profile from the N-body simulations (black line) as well as its parametrization with the MGE Gaussians (green line). Also shown is the resulting model compared to the data (lower left panel). Model provides a worse fit then the model with a constant $M/L_V$ profile. This comes probably from overestimating the $M/L_V$ profile in the outer regions of the cluster.

Both methods of deriving an $M/L_V$ profile show similar results. Figure \ref{fig:mod_axi} in the upper panel shows the Jeans model for the major axis, minor axis and circular averaged on top of our data points. The model of the major axis seems to reproduce the data best. This might be due to the fact that the data set of the Fabry-Perot observations is asymmetric and spatially biased towards the major axis. The right panel of the figure shows the deprojected $M/L$ profile from the N-body simulations (black line) together with the fit of the $M/L$ profile from the Jeans model (red line). The $M/L$ profile in the lower panel of Figure \ref{fig:mod_axi} has an interesting shape. The steep rise at the center implies a high concentration of stellar remnants and therefore an advanced stage of mass segregation. Beyond the half-light radius ($\sim 48''$) the $M/L$ profile rises again, which can also be explained by the process of mass segregation. Low mass stars move towards the outer regions while the cluster evolves. That explains the higher $M/L$ ratio in the outskirts of the cluster.  

The next step in terms of modeling is to include a central black hole in our Jeans models and to test if we obtain a better fit to our data. Figure \ref{fig:mod_bh} shows the result of these models. For both cases, with and without a $M/L_V$ profile, we compute models with black-hole masses between $M_{BH} = 0 $ and $M_{BH}=8\times 10^3 \ M_{\odot}$. The black solid line shows the best fit of the model with a zero mass black hole. Higher black-hole masses predict higher central V$_{RMS}$ than seen in our data. The $\chi^2$ curves in the right panel of the figure shows this result. The blue shaded areas define the $1 \sigma$ limit of the best fit. For the model with the constant $M/L_V$ profile, no black hole is needed in order to reproduce the data with a $1 \sigma$ uncertainty of $M_{\bullet} = 1 \times 10^3 M_{\odot}$. The second model, however, exhibits a flatter profile and therefore is better fit by a model with a non-zero black-hole mass. The best fit model returns a black-hole mass of $M_{\bullet} = 4 \times 10^3 M_{\odot}$ and a $1 \sigma$ upper limit of $M_{\bullet} \sim 6 \times 10^3 M_{\odot}$. Also the total $M/L_V$ values of both models differ. The model with the constant $M/L_V$ results in a total value of $M/L_V = 1.6 M_{\odot}/L_{\odot}$ while the model with the $M/L_V$ increasing in the outskirts of the cluster predicts a $M/L_V$ of $\sim 2.1 M_{\odot}/L_{\odot}$.

We run Monte Carlo simulations on both the surface-brightness profile and the velocity-dispersion profile in order to get an estimate of how much the individual errors from the two profiles influence the result. We find that by only changing the inner surface-brightness profile, 90~\% of the cases result in the same black-hole mass as the fit with the original light profile. This indicates that the error on the black-hole mass is only dependent on the uncertainties of the velocity-dispersion profile. We therefore adopt the $95 \%$ confidence upper limit of $M_{\bullet} < 1 \times 10^4 M_{\odot}$ of the model with the varying $M/L_V$ profile. We note that the model with the constant $M/L_V$ fits the data slightly better and results in a lower upper limit of $M_{\bullet} < 3 \times 10^3 M_{\odot}$. However, a constant $M/L_V$ profile is very unrealistic for a globular cluster like NGC 2808 and therefore the results from the model with the $M/L_V$ profile more reliable. From the Monte Carlo simulations we also derive a $1 \sigma$ error on the total mass and the global mass to light ratio of $\sim 10~\%$.


\section{Summary and Conclusions} \label{con}
                                   

We examine the central kinematics of the globular cluster NGC 2808 in order to constrain the mass of a possible intermediate-mass black hole at its center. With a set of HST images, the photometric center of the cluster is determined. Furthermore, a color magnitude diagram as well as a surface brightness profile, built from a combination of star counts and integrated light, are produced. The spectra from the VLT ground-based integral-field unit ARGUS are reduced and analyzed in order to create a velocity map and a velocity-dispersion profile. We derive a velocity-dispersion profile by summing all spectra into radial bins and applying a penalized pixel fitting method. In addition, we use radial velocities measured with the Rutgers Fabry Perot by Gebhardt et al. (2012, in preparation) to derive the velocity-dispersion profile in the outer regions.

We compare the data to spherical and axisymmetric isotropic Jeans models using different $M/L_V$ profiles and black-hole masses. We include a radius dependent $M/L_V$ profile obtained from N-body simulations in order to account for mass segregation in the cluster. This does not improve the fit of the velocity dispersion profile, but still coincides with the data within the error bars. The best fitting Jeans model is the axisymmetric case with a constant $M/L_V$ profile and no black hole. The $95 \%$ error of that fit predicts an upper limit of $M_{\bullet} = 3 \times 10^3 M_{\odot}$ on the black-hole mass. However, assuming a non constant $M/L_V$ profile the model upper limit on the black-hole mass increases to $M_{\bullet} = 1 \times 10^4 M_{\odot}$.

Our result on the upper limit of the black-hole mass in NGC 2808 is higher with the results of the radio observations of \cite{maccarone_2008} and their upper limit of 370 - 2100 $M_{\odot}$. With the uncertainties in gas content and accretion rates, however it is plausible for the limit derived by the radio observations to increase. We stress that the result is dependent on the choice of the $M/L_V$ profile and needs to be treated carefully since different $M/L_V$ profiles bring different results on black-hole mass and total mass of the cluster. 

Our derived mass to light ratio of $M/L_V = (2.1 \pm 0.2) \ M_{\odot}/L_{\odot}$ is higher than the M/L of $\sim 1.3$ derived by \cite{McLaughlin_2005}. This results from the fact that our total luminosity is lower than the one derived by \cite{harris_1996} by $\sim 22 \%$ which results from the slightly steeper drop of the surface brightness profile compared to the profile obtained by \cite{trager_1995} in the outskirts of the cluster. Also the total mass of $M_{TOT} = (8.2 \pm 0.8) \times 10^5  M_{\odot}$ is higher than the total mass derived with the values of \cite{harris_1996} and \cite{McLaughlin_2005} of $M_{TOT} \sim 6.4 \times 10^5  M_{\odot}$ which results from the shape of our $M/L_V$ profile.

So far, Jeans models allow us a crude first guess on the dynamic state of a globular cluster. Nevertheless, the result of the black hole-mass depends strongly on the $M/L_V$ profile used. We find that using different profiles with lower M/L values at the center but still high values in the outskirts result in models which fit the data with a black-hole mass up to 5000 $M_{\odot}$. However, due to the high degeneracy we encounter when fitting the M/L profile with Jeans models, we have to choose the profile which is derived from N-body simulations. Therefore, it is crucial to run specific N-body simulations for all our globular cluster in order to get an accurate $M/L_V$ profile and constraints on the anisotropy and mass segregation. This is an important factor especially for mass segregated clusters such as NGC 2808. 

The dynamical models presented here include Jeans isotropic modeling and comparison with N-body simulations. We do not present results from orbit-based models. These axisymmetric models are significantly more general than the isotropic models since they have no assumption about the velocity anisotropy. They are also more general than the N-body simulations, since the N-body models rely on a limited set of initial conditions; the orbit models encompass all available phase space configurations, at the expense of producing a dynamical model that does not take the evolutionary processes into account. Thus, the orbit models will provide larger uncertainties, and hence a large upper limit on the black hole mass. For this analysis, however, we rely on the N-body simulations which should be a fair representation of the current dynamical state of the cluster.

The study of black holes in globular clusters has drawn the attention of the astronomy community. Not only radial velocities are observed and analyzed, also, observations and analysis of proper motions for many clusters are in progress. A desired future project would be the combination of all these data sets and a detailed analysis via N-body and orbit based models which would allow a deeper insight into the dynamics of globular clusters, revealing their hidden secrets. 

\begin{acknowledgements}
This research was supported by the DFG cluster of excellence Origin and Structure of the Universe (www.universe-cluster.de). H.B. acknowledges support from the Australian Research Council through Future Fellowship grant FT0991052. We thank an anonymous referee whose fast comments and suggestions have helped a lot to improve this paper. 
\end{acknowledgements}

\bibliographystyle{aa}
\bibliography{ref}

\end{document}